\documentclass[prd, amsfonts, twocolumn, nofootinbib, showpacs]{revtex4}

\usepackage{graphicx, epsfig}
\usepackage{color}
\usepackage{amsmath}
\newcommand{\be}{\begin{equation}}
\newcommand{\ee}{\end{equation}}
\newcommand{\bea}{\begin{eqnarray}}
\newcommand{\eea}{\end{eqnarray}}

\newcommand{\gapp}{\mathrel{\raise.3ex\hbox{$>$}\mkern-14mu
\lower0.6ex\hbox{$\sim$}}}
\newcommand{\lapp}{\mathrel{\raise.3ex\hbox{$<$}\mkern-14mu
\lower0.6ex\hbox{$\sim$}}}
\def\bbox{{\,\lower0.9pt\vbox{\hrule \hbox{\vrule height 0.2 cm
\hskip 0.2 cm \vrule  height 0.2 cm}\hrule}\,}}

\begin{document}
\title{Origin of the tail in Green's functions in odd dimensional space-times}
\author{De-Chang Dai$^1$ and Dejan Stojkovic$^2$}
\affiliation{$^1$ Institute of Natural Sciences and INPAC, Department of Physics,
Shanghai Jiao Tong University, Shanghai 200240, China}
\affiliation{$^2$ HEPCOS, Department of Physics, SUNY at Buffalo, Buffalo, NY 14260-1500}

 %%%%%%%%%%%%%%%%%%%%%%%%%%%%%%%%%%%%%%%%%%%%%%%%%%%%%%%

\begin{abstract}
\widetext
It is well known that the scalar field Green's function in odd dimensions has a tail, i.e. a non-zero support inside the light cone, which in turn implies that the Huygens' principle is violated.  However, the reason behind this behavior is still not quite clear.  In this paper we shed more light on the physical origin of the tail by regularizing the term which is usually ignored in the literature since it vanishes due to the action of the delta function. With this extra term the Green's function does not satisfy the source-free wave equation (in the region outside of the source). We show that this term corresponds to a charge imprinted on the light cone shell.  Unlike the vector field charge, a moving scalar field charge is not Lorentz invariant and is contracted by a $\sqrt{1-v^2}$ factor. If a scalar charge is moving at the speed of light, it appears to be zero in the static (with respect to the original physical charge) observer's frame. However, the field it sources is not entirely on the light cone. Thus, it is likely that this hidden charge sources the mysterious tail in odd dimensions.
\end{abstract}

%%%%%%%%%%%%%%%%%%%%%%%%%%%%%%%%%%%%%%%%%%%%%%%%%%

%\pacs{}
\maketitle

\section{Introduction}

While studying the propagation of light in ether, around 1690, Huygens concluded that perturbations produced by a light source placed at $(r=0,t=0)$ propagate away in the form of spherical waves with uniform velocity. Every spatial point reached by the wave front behaves as a new source and produces a new spherical wave. As a result, an entire envelope of the spherical waves in space is formed \cite{Huygens}. This is the historic idea of the Huygens' principle, which proved very useful in our studies of wave propagation phenomena and interference (including the famous double slit experiment).

Then, around 1900, Hadamard came to conclusion that the Huygens' principle in odd-dimensional space-times is violated \cite{Hadamard}. Since Huygens' principle appears to be true only in even dimensional space-times, Paul Ehrenfest argued that the special role of $(3+1)$-dimensional d'Alembert wave equation gives an indication why our world is $(3+1)$-dimensional \cite{Ehrenfest}.

Since Huygens' principle is very closely related to the method of Green's functions, violation of Huygens's principle in odd dimensional space-times means the odd dimensional Green functions are also very peculiar. Physically, a delta function source on the right hand side of d'Alembert wave equation, i.e. $\delta (t-t_0)\delta (\vec{r}-\vec{r}_0)$, means that a source is localized both in space and time. So, a source (i.e. charge) appears and disappears at $t=0$, the field responds and starts propagating away from the source. If the Huygens' principle is violated, the field will not propagate at a constant velocity for all of the frequency modes (discussion about the related issues in curved space-time can be found in \cite{Dai:2012ni,Chu:2011ip} and references therein). Phenomena will then not depend only on an instantaneous perturbation,  but on the whole past history due to the infinite tail that the source leaves in odd dimensions. This fact will make the study of radiation and interactions very inconvenient \cite{Murakami:2002jy,Caldwell:1993xw,Dahl,Poisson:2003nc,Frolov:2003mc}.

The infinite tail in odd dimensions can be nicely illustrated using dimensional reduction \cite{ajp}. While the dimensional reduction is technically useful, it does not shed any light on a problem in a space-time with the fixed number of dimensions. In particular, dimensional reduction procedure uses the charge located in extra dimensions perpendicular to the original space to explain the tail that appears in the original space. In the absence of extra dimensions, this extended charge must somehow be hidden in the original space. The only place where we can hide is the light cone.

The goal of this paper is to find the physical source of the violation of the Huygens' principle so that it can be treated properly. We first observe that the odd dimensional Green's functions are discontinuous at the light cone, which implies the existence of the charge on the light cone shell. If we integrate the volume near the light cone, we will find that the total charge is zero. However, a charge located on the light cone is moving with the speed of light, so the Lorentz contraction will reduce the total charge to zero for an observer in the original frame where the source charge was placed at the origin. But the charge actually exists, what we can verify by direct calculations. In particular, we can regularize our calculations by considering a shell which is moving at some velocity close to the speed of light and then taking the speed of light limit.

It is important not to confuse the scalar field charge with a vector field charge in this case. A vector field charge, e.g. electric charge, is a Lorentz invariant quantity, while a scalar field charge is not. What is invariant is the scalar field charge density. But the volume is Lorentz contracted by a boost, so the charge must be Lorentz contracted too \cite{Adelberger:1992ph,Ren:1993bs}.

The outline of the paper is as follows. We will first introduce the basic idea of the Green's functions method and violation of the Huygens's principle. Then we will explicitly show that the scalar field charge is not a Lorentz invariant quantity. This will explain why we do not see the extra light-cone charge in the static observer's frame. We will then show that  this charge is not zero in the co-moving frame on the light cone. This indicates that the source of the tail of the Green's function in odd dimensional space-times (or the violation of the Huygens' principle) can be interpreted as the charge on the light cone shell. We will then study a known charge distribution and calculate the magnitude of the hidden charge for it.

\section{ Green's functions and violation of the Huygens principle}

Consider a scalar field, $\psi(t,\vec{r})$, wave equation in a $(d+1)$-dimensional space-time  with a scalar field charge distribution $f(t,\vec{r})$

\begin{equation}
\label{wave_1}
\partial_t^2 \psi-\sum_i^d\partial_{x_i} ^2\psi=f(t,\vec{r})
\end{equation}
Here, $\vec{r}=(x_1,x_2,...,x_d)$. This equation can be solved if one can find the solution for the Green's function $G(t-t_0,\vec{r}-\vec{r_0})$ in
\begin{equation}
\label{green_o}
\partial_t^2 G-\sum_i^d\partial_{x_i} ^2 G=\delta(t-t_0)\delta(\vec{r}-\vec{r_0})
\end{equation}
The solution to Eq.~(\ref{wave_1}) is then obtained directly from
\begin{equation}
\label{green}
\psi=\int G(t-t_0,\vec{r}-\vec{r_0})f(t_0,\vec{r_0})dt_0d\vec{r_0}
\end{equation}

The Green's functions for the scalar field are well known \cite{Hassani,Jackson}. In even dimensional space-times (i.e. $d$ is odd), they are
\begin{eqnarray}
_eG^{d+1}_{ret}(t,r)&=&\frac{1}{4\pi}\Big(-\frac{1}{2\pi r}\partial_r\Big)^{(d-3)/2}\Big(\frac{\delta (t-r)}{r}\Big)\\
_eG^{d+1}_{adv}(t,r)&=&\frac{1}{4\pi}\Big(-\frac{1}{2\pi r}\partial_r\Big)^{(d-3)/2}\Big(\frac{\delta (t+r)}{r}\Big)
\end{eqnarray}
where $_eG_{ret}^{d+1}$ is the retarded, while $_eG_{adv}^{d+1}$ is the advanced solution. Take the $(3+1)$-dimensional retarded solution as an example
\begin{equation}
_eG^{3+1}_{ret}(t,r)=\frac{1}{4\pi}\frac{\delta (t-r)}{r} .
\end{equation}
Since the source term for the Green's function is $\delta(t)\delta(\vec{r})$, the charge suddenly appears and disappears at $(t,\vec{r}) =(0,\vec{0})$. The scalar field will then respond to this perturbation and start to propagate outward from the origin. Since the source is absent for $t>0$, the propagation of the field should be source-free for all $t>0$. The term $\delta(t-r)$ in the solution implies that the propagation velocity is exactly the speed of light and is constant (not position dependent). Since all the (frequency) modes of the signal are propagating at the same speed, the Huygens's principle will be valid. All the other even dimensional Green's functions in $d > 3$ include the $\delta(t-r)$ term and its derivatives. Therefore the signals in these cases also propagate at the speed of light and the Huygens's principle is preserved.

\begin{figure}[h]
\centering
\includegraphics[width=8cm]{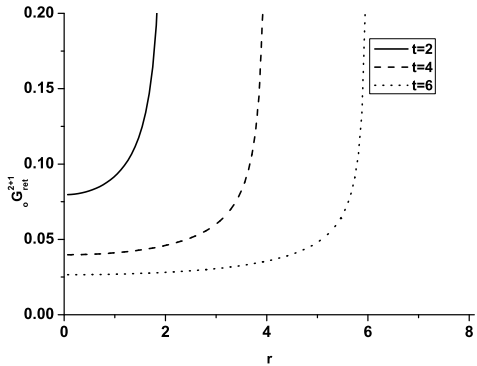}
\caption{This figure shows $(2+1)$-dimensional retarded Green's function at $t=2$, $t=4$ and $t=6$. Unlike even dimensional space, the green function is not zero in the $t>r$ region of the space-time, but it has an infinite tail. This feature ruins the Huygens' principle.}
    \label{2+1}
\end{figure}

Although the Huygens' principle is very intuitive, and one would expect it to hold in general, it happens somehow that it is not preserved in odd-dimensional space-times. This can be seen directly from the odd dimensional Green's functions
\begin{eqnarray}
\label{odd-1}
_oG^{d+1}_{ret}(r,t)&=&\frac{\theta(t)}{2\pi}\Big(-\frac{1}{2\pi r}\partial_r\Big)^{d/2-1}\Big(\frac{\theta(t-r)}{\sqrt{t^2-r^2}}\Big)\\
\label{odd-2}
_oG^{d+1}_{adv}(r,t)&=&\frac{\theta(-t)}{2\pi}\Big(-\frac{1}{2\pi r}\partial_r\Big)^{d/2-1}\Big(\frac{\theta(-t-r)}{\sqrt{t^2-r^2}}\Big).
\end{eqnarray}
where $_oG_{ret}^{d+1}$ is the retarded, while $_oG_{adv}^{d+1}$ is the advanced solution. It is important to note that the Heaviside step function $\theta(t-r)$ is now present in the solution, which was not the case in even-dimensions. This implies that some modes travel slower than the speed of light. Take the $(2+1)$-dimensional retarded Green's function as an example
\begin{equation}\label{2d}
_oG^{2+1}_{ret}(r,t)=\frac{1}{2\pi}\frac{\theta(t-r)}{\sqrt{t^2-r^2}}
\end{equation}
Fig.~\ref{2+1} shows that the Green's function is not zero in the $t>r$ region of the space-time. The tail is the evidence that some modes of the wave travel at speeds slower than the speed of light. The tail never becomes zero at the source location, though the source disappeared after the initial appearance at $t=0$. It therefore appears that the Huygens' principle can not survive in an odd dimensional space-time.

It is somewhat surprising that the Huygens' principle is a dimensionally dependent effect, since it contradicts our physical intuition. Any influence inside the light cone must have a corresponding source, which is clearly not the source at the origin.  In the rest of the paper we will show that the existence of the tail is connected with an existence of additional charge on the light cone.

\section{Dissecting the Green's function in $(2+1)$-dimensional space-time}

We begin with observation that the Green's function in Eq.~(\ref{2d}) is discontinuous on the light cone $t=r$. It is $1/\sqrt{t^2-r^2}$ inside the light cone but it is zero outside the light cone. Obviously there is a discontinuity exactly on the light cone. From Gauss' law one expects to find the charge at the light cone which causes this discontinuity. Since this discontinuous field distribution moves at the speed of the light, we expect the charge also to move at the speed of the light. In this section we will show that explicitly.

Although both odd- and even-dimensional Green's functions come from solving Eq.~(\ref{green_o}), there is a fundamental difference between them. If we substitute the solution to the even-dimensional Green's function back into Eq.~(\ref{green_o}), we can verify that the original equation is indeed satisfied. On the other hand, if we do the same for the     odd-dimensional Green's functions, we will find that Eq.~(\ref{green_o}) is not precisely satisfied. Take again a $(2+1)$-dimensional retarded Green's function in Eq.~(\ref{2d}) as an example. Substituting it back into Eq.~(\ref{green_o}) one finds
\begin{eqnarray} \label{main}
\partial_t^2 {_oG^{2+1}_{ret}}&-&\sum_i^2\partial_{x_i} ^2 {_oG^{2+1}_{ret}}=\delta(t)\delta(\vec{r})\nonumber\\
&+&\frac{1}{2\pi}\frac{\sqrt{t^2-r^2}}{r(r+t)^2}\delta(t-r)
\end{eqnarray}
Obviously, there is an extra term on the right hand side. Formally, this extra term becomes zero upon the action of the delta function since it contains $\sqrt{t^2-r^2}$.
Since the delta function expressions make clear physical sense only under the integral, and integration in this case will make this term vanish, one option is to ignore it as it is usually done in the literature.
However, if we want to shed more light to the problem, we could try to regularize this term and see what it corresponds to.  For example, we can rewrite this term as
\begin{eqnarray}\label{rt}
&&\frac{1}{2\pi}\frac{\sqrt{t^2-r^2}}{r(r+t)^2}\delta(t-r)=\frac{1}{2\pi}\frac{t\sqrt{1-v^2}}{r(r+t)^2}\delta(t-r)|_{v=\frac{r}{t}}\nonumber\\
&=&\lim_{v\rightarrow 1} \frac{1}{2\pi}\frac{t\sqrt{1-v^2}}{r(r+t)^2}\delta(vt-r)\nonumber\\
&=&\lim_{v\rightarrow 1} \frac{1}{2\pi}\frac{t}{r(r+t)^2}\delta\left(\frac{vt-r}{\sqrt{1-v^2}}\right) .
\end{eqnarray}
Here, we replaced  $\sqrt{t^2-r^2}$ with $t\sqrt{1-v^2}$, where $v=r/t$. Since $\delta(t-r)$ implies $v=1$, we can add $v$ in the $\delta(vt-r)$ without affecting the value of the term. In this new form, $\sqrt{t^2-r^2}$ becomes the relativistic boost factor, $\sqrt{1-v^2}$. We will then infer that there is a net charge on the $t=r$ shell, but because the shell moves at the speed of light, the charge is Lorentz contracted and becomes zero in the static observer's frame. In particular, the second line in Eq.~(\ref{rt}) will have the form of the charge as in Eq.~(\ref{c}), while the third line in Eq.~(\ref{rt}) will have the form of the charge distribution as in Eq.~(\ref{cd}).

One important fact is that the scalar charge, unlike a vector field charge, is not Lorentz invariant. What is Lorentz invariant is the scalar charge density, but since the volume experiences Lorentz contraction the scalar charge must contain a factor of $\sqrt{1-v^2}$. We can illustrate this by looking at a static charge solution in $(3+1)$-dimensions. The solution corresponding to the static scalar charge $q$ at $\vec{r}=\vec{0}$ is
\begin{equation}
\psi=\frac{q}{4\pi r}  .
\end{equation}
If we substitute this solution into  Eq.~(\ref{wave_1}) we find the charge distribution
\begin{equation}
f(t,\vec{r})=q\delta(\vec{r})  .
\end{equation}
The total charge is
\begin{equation}
\int f(t,\vec{r})d\vec{r}=q  .
\end{equation}
We now boost to a new referent system
\begin{eqnarray}
t&=&\frac{t'-vx'}{\sqrt{1-v^2}}\\
x&=&\frac{x'-vt'}{\sqrt{1-v^2}}\\
y&=&y'\\
z&=&z'
\end{eqnarray}
Then the scalar field in new coordinates becomes
\begin{equation}
\psi'=\frac{q}{4\pi \sqrt{\left(\frac{x'-vt'}{\sqrt{1-v^2}}\right)^2+y'{^2}+z'{^2}}}
\end{equation}
We can calculate the corresponding charge density again by plugging this form into Eq.~(\ref{wave_1}). The charge distribution is
\begin{equation}\label{cd}
f'(t',\vec{r'})=q\delta \left(\frac{x'-vt'}{\sqrt{1-v^2}}\right)\delta(y')\delta(z')
\end{equation}
Therefore, the total scalar charge in the new system is
\begin{equation}\label{c}
q'=\int f'(t',\vec{r'})d\vec{r}=q\sqrt{1-v^2}
\end{equation}
Obviously, the magnitude of the charge is reduced by a factor of $\sqrt{1-v^2}$. As a consequence, if the scalar charge is moving at the speed of light in some referent system, then this charge is not visible in that referent system.
Therefore, if the $(2+1)$-dimensional Green's function in question does include a charge which moves at the speed of light, then the magnitude of that charge will be zero even though the charge does exist on the light cone shell.

We can now study some known scalar charge distribution and find out whether it is possible to identify the exact form of the hidden charge for such distribution. We try the following charge distribution in a $(2+1)$-dimensional space-time
\begin{equation}
\label{charge-1}
f(t,\vec{r})=q\theta(t)\delta(\vec{r})  .
\end{equation}
This form describes a charge, $q$, that appears at $(t,\vec{r})=(0,\vec{0})$ and stays there forever. The corresponding field distribution can be obtained from Eq.~(\ref{green})
\begin{eqnarray}
\psi (t,r) &=&\int {_o}G^{2+1}_{ret}(t-t_0,\vec{r}-\vec{r_0})f(t_0,\vec{r_0})dt_0d\vec{r_0}\nonumber\\
&=&-\int_0^t \frac{q}{2\pi}\frac{\theta(t'-r)}{\sqrt{t'{^2}-r^2}}dt'\nonumber\\
\label{2+1-solution}
&=&-\frac{q}{2\pi}\ln\Big(\frac{t+\sqrt{t^2-r^2}}{r}\Big)\theta(t-r)
\end{eqnarray}
If we substitute this solution into Eq.~(\ref{wave_1}), we will find that the charge density in the neighborhood $t=r$ is
\begin{equation}
\rho_s(t,\vec{r})=\frac{1}{2\pi}\Big(\frac{-\ln(\frac{t+\sqrt{t^2-r^2}}{r})}{r}+\frac{2\sqrt{t^2-r^2}}{r(r+t)}\Big)\delta(t-r) .
\end{equation}
The total charge on this shell is found by integrating the volume in vicinity of the light cone, i.e. from $r=t-\delta$ to $r=t+\delta$ for some vanishing $\delta$
\begin{equation}
q_s=\int_{r^-}^{r^+} \rho_s(t,\vec{r})2\pi rdr =0.
\end{equation}
 The magnitude of the scalar charge is exactly zero. However, as we already explained, this is because of the Lorentz contraction. We can actually verify that the total charge is not really zero. Since the $t=r$ shell is moving at the speed of light, we cannot find the charge by a simple boost.
 Instead, we "regularize" the solution by inserting the parameter $v$, but will be interested in the $v \rightarrow 1$ limit at the end. This limit will recover the original solution (and the distribution), but will also shed the light on the hidden charge. The hidden charge will be then found by boosting the distribution with a finite $v$, calculating the charge in the shell's rest frame, and then taking the limit of $v\rightarrow 1$. So, according to this, we first modify Eq.~(\ref{2+1-solution}) to
\begin{eqnarray}
\psi_v=-\frac{q}{2\pi}\ln\Big(\frac{t+\sqrt{t^2-r^2}}{r}\Big)\theta(vt-r) .
\end{eqnarray}
This equation describes a shell moving at some finite speed $v$. This shell must contain some charge, because some of the field is screened out. We plug this equation into Eq.~(\ref{wave_1}) and find the charge distribution near the shell $r=vt$ as
\begin{eqnarray}
&&\rho_{sv}(t,\vec{r})=\frac{q}{2\pi}\Bigg(\ln\Big(\frac{t+\sqrt{t^2-r^2}}{r}\Big)(v^2-1)\delta'(vt-r)\nonumber\\
&&+\bigg(-\frac{\ln(\frac{t+\sqrt{t^2-r^2}}{r})}{r}+\frac{2(t-vr)}{r\sqrt{t^2-r^2}}\bigg)\delta(vt-r)\Bigg)
\end{eqnarray}
We can find the magnitude of the charge on the shell by directly integrating the volume from $r=vt-\delta$ to $r=vt+\delta$, for small $\delta$
\begin{eqnarray} \label{qsv}
&q_{sv}&=\int_{vt-\delta}^{vt+\delta}\rho_{sv}(t,\vec{r})2\pi rdr\nonumber\\
&=&q\left[\left(-2+v^2\right)\ln \left(\frac{1+\sqrt{1-v^2}}{v}\right)+3\sqrt{1-v^2}\right]
\end{eqnarray}

As expected, for $v\sim 1$, the charge is proportional to the boost factor, i.e. $q_{sv}\propto \sqrt{1-v^2}$. But we need to calculate the charge in the shell's rest frame. Since the shell is moving with velocity $v$ in $\hat{r}$ direction, the normal vector perpendicular to the shell's hypersurface is $(n^t,n^r)=(\frac{v}{\sqrt{1-v^2}},\frac{1}{\sqrt{1-v^2}})$. Therefore when the shell is at $(t,r)=(t_0,vt_0)$, the perpendicular direction can be written as $(l^t,l^r)=(t_0+n^t\epsilon,vt_0+n^r\epsilon)$. Then the charge in the shell's rest frame is
\begin{eqnarray}
&q_{nsv}&=\int_{vt_0-\delta}^{vt_0+\delta}\rho_{sv}(t,\vec{r})2\pi rdl\nonumber\\
&=&\int_{vt_0-\delta}^{vt_0+\delta}\rho_{sv}(t_0+n^t\epsilon,vt_0+n^r\epsilon)2\pi rd\epsilon\nonumber\\
&=&q\left[\frac{-2}{\sqrt{1-v^2}}\ln \left(\frac{1+\sqrt{1-v^2}}{v}\right)+3\right]
\end{eqnarray}
Therefore, the magnitude of the charge which moves with the speed of light is
\begin{equation}
Q=\lim_{v\rightarrow 1}q_{nsv}=q
\end{equation}
This proves that there is a non-zero charge on the $r=t$ shell. It appears to be zero for a static observer (with respect to the original physical charge) only because the shell is moving at the speed of light.
So in contrary to the naive expectations, the propagation of the field is not source-free outside the origin. Information about the original charge somehow remains embedded into the light cone shell in odd dimensions.

\section{$(4+1)$ and higher odd-dimensional Green's functions }
\label{41}

As noticed in \cite{Aharonovicha}, there seems to be some ambiguity in the literature when higher odd-dimensional Green's functions are discussed. For example we quote results from  \cite{Andrei,Vladimirov}
\begin{eqnarray}
_{no}G^{d+1}_{ret}(r,t)&=&(-1)^{\frac{d-2}{2}}\frac{\Gamma(\frac{d-1}{2})}{2\pi^{\frac{d+1}{2}}}\frac{\theta(t-r)}{(t^2-r^2)^{\frac{d-1}{2}}}\\
_{no}G^{d+1}_{adv}(r,t)&=&(-1)^{\frac{d-2}{2}}\frac{\Gamma(\frac{d-1}{2})}{2\pi^{\frac{d+1}{2}}}\frac{\theta(-t-r)}{(t^2-r^2)^{\frac{d-1}{2}}}
\end{eqnarray}
These Green's functions are different from  the ones we presented in Eqs.~(\ref{odd-1}) and (\ref{odd-2}), but they differ only on the light cone. We look at the $(4+1)$-dimensional Green function as an example. The difference is
\begin{equation}
_{o}G^{4+1}_{ret}(r,t)-_{no}G^{4+1}_{ret}(r,t)=\frac{1}{4\pi^2}\frac{\delta(t-r)}{r\sqrt{t^2-r^2}}
\end{equation}
Since the difference is only on the light cone, it is usually ignored. However, this difference will generate infinite amount of charge on the light cone and make the calculations significantly more difficult. To illustrate this, consider a source
\begin{equation}
\label{charge}
f(t,\vec{r})=\theta(t)\delta(\vec{r})
\end{equation}
When we substitute this source in Eq.~(\ref{green_o}), we get the field
\begin{eqnarray}
\psi_o (t,r) &=& \int _0^{t}  {_o G^{4+1}_{ret}}(t_0,r) dt_0\\
\label{4+1psi}
&=&\frac{1}{4\pi^2 r^2}\frac{t}{\sqrt{t^2-r^2}} \label{lcs}
\end{eqnarray}
In the limit of $t\rightarrow \infty$, we recover the static case, $\psi=\frac{1}{4\pi^2 r^2}$. In the limit of $r\rightarrow 0$, we also recover $\psi=\frac{1}{4\pi^2 r^2}$. If we now apply the Gaussian surface integral we will get the correct charge at $r=0$. However, if we apply the same source (\ref{charge}) to the $_{no} G^{4+1}_{ret}$, then the field will become
\begin{eqnarray}
\psi_{no}(t,r)&=&-\int_r^{t} dt_0\frac{1}{4\pi^2 (t^2-r^2)^{3/2}}\nonumber\\
&=&\frac{1}{4\pi^2 r^2}\Big(\frac{t}{\sqrt{t^2-r^2}}-\frac{t_0}{\sqrt{t_0^2-r^2}}\Big)|_{t_0=r}
\end{eqnarray}
This expression is divergent at all times. Therefore it can not give the correct potential, unless one finds a way to regularize the divergent terms. This fact indicates that the solution $_{no} G^{4+1}_{ret}$ can not be the correct one. The other way to see that $_{no} G^{4+1}_{ret}$ is not the correct solution is to substitute $_{no} G^{4+1}_{ret}$ into $(4+1)$-dimensional spherical wave equation (\ref{green_o})
\begin{eqnarray}
&&\frac{\partial^2}{\partial t^2} {_{no} G^{4+1}_{ret}}-\frac{1}{r^3}\frac{\partial}{\partial r}\Big(r^3 \frac{\partial}{\partial r} {_{no} G^{4+1}_{ret}}\Big)\nonumber\\
&&=-\frac{3}{4\pi^2}\frac{\delta(t-r)}{\sqrt{t^2-r^2}(r+t)^2r}
\end{eqnarray}
This equation is obviously non-vanishing on the light cone. Therefore it can not be the right Green's function. However, the same equation also tells us that the tail in ${_{no}} G^{4+1}_{ret}$ is generated by infinite amount of charge on the light cone. Since the tail is generated by the charge at the light cone instead of the charge at the center, one can not conclude that the information of a perturbation can last forever. Since ${_{no}} G^{4+1}_{ret}$ and ${_{o}} G^{4+1}_{ret}$ have the same tail, this implies that the tail of ${_{o}} G^{4+1}_{ret}$ is also from the charge on the light cone.

The solution ${_{o}} G^{4+1}_{ret}$ also has some charge on the light cone, but it is finite, just like in the case of $(2+1)$-dimensional Green's function. Indeed, we can apply $_{o} G^{4+1}_{ret}$ to the $(4+1)$-dimensional spherical wave equation and find
\begin{eqnarray}
&&\frac{\partial^2}{\partial t^2} {_{o} G^{4+1}_{ret}}-\frac{1}{r^3}\frac{\partial}{\partial r}\Big(r^3 \frac{\partial}{\partial r} {_{o} G^{4+1}_{ret}}\Big)\nonumber\\
&&=-\frac{1}{4\pi^2}\left[\frac{2r^2-2rt-t^2}{\sqrt{t^2-r^2}(r+t)^2r^3}\delta(t-r) \right. \nonumber\\
&&\left. -\frac{t-r}{\sqrt{t^2-r^2}(r+t)r^2}\delta '(t-r)\right]
\end{eqnarray}
The solution does not accumulate infinite charge on the light cone. Careful calculations show that the charge density is $\propto \sqrt{t^2-r^2} \delta(t-r)$, so this factor is just a boost effect, similar to the one we had in $(2+1)$ dimensions. Following the procedure we developed for $(2+1)$ dimensions, one can remove the boost effect and find out the true total charge. Therefore, the tail in higher odd dimensions is perhaps also generated by the charge on the light cone, instead of the charge at the center.

\section{Conclusions}

We studied the Green's functions for a scalar field in odd-dimensional space-times in order to understand the appearance of the tail, i.e. the fact that the Green's function has a non-zero support inside the light cone. Observation that the Green's function in Eq.~(\ref{2d}) is discontinuous on the light cone $t=r$ (it is $\sim 1/\sqrt{t^2-r^2}$ inside the light cone but it is zero outside the light cone, indicates (by Gauss' Law) that there is a charge on the light cone which causes this discontinuity. Indeed, we demonstrated that the Green's function in odd dimensions is not exactly a charge-free solution (in the region away from the origin). In doing so we diverged from the standard procedure in dealing with distributions to ignore the term which will vanish upon integration because of the presence of the delta function. Instead we ``regularized" this extra term to show that it represents the non-zero charge on the light cone. This charge is most likely the source of the tail at $t>r$. The total charge appears to be zero only because the light cone charge is moving at the speed of light and experiences the Lorentz contraction. Thus, the total charge is vanishing it the static observer's frame, but it still affects the field inside its causal past. As an illustration, we studied a known scalar charge distribution in Eq.~(\ref{charge-1}) and found the exact form of the hidden charge for it.

In even dimensions, there is no extra term on the right hand side of Eq.~(\ref{main}). Therefore there is no light-cone charge as an extra source in even dimensions, which is required for the self-consistency of our method.

The case of the vector field charge is more complicated. Naively, the vector Green's function $G_{\mu \nu}(x; x')$ in the
Lorenz gauge is proportional to the scalar Green's function $G (x; x')$, i.e. $G_{\mu \nu}(x; x') = \eta_{\mu \nu} G(x; x')$. However, it is important note that the physical vector charge is Lorenz invariant, so it can't be boosted to zero. When boosting a vector charge one needs to include both the density $\rho$ and current $\vec{J}$, in contrast with the scalar field where only $\rho$ is boosted.

It appears that that behavior of the fields in even and odd dimensions is somehow fundamentally different. The appearance of the tail in odd dimensions and absence of it in even dimensions is just one example.  The other example is the question of the non-existence/existence of the Randall-Sundrum black holes is various dimensions. While there is a solution  in the $(2+1+1)$ case ($(2+1)$-dimensional space-time plus one extra dimension), there is still no satisfactory solution in the $3+1+1$ case \cite{Tanaka:2002rb,Emparan:2002px}
(see \cite{Figueras:2011gd,Abdolrahimi:2012pb,Dai:2010jx} for solutions in some limiting cases), and the existence and non-existence alternate from even to odd dimensions. Again, odd and even dimensions behave completely differently. In the case of Randall-Sundrum black holes, holography - the fact that information about the bulk is also encoded on the boundary - is often invoked as an explanation. We may speculate at the end that the current case with Green's functions also looks like holography - information about the charge at the origin is encoded on the light-cone shell. The point is that the boundary does not have to be a true physical boundary, it could be an event horizon of the black hole, or a cosmological horizon. It is intriguing that the boundary may also be a light-cone shell.

\acknowledgments{
This work was partially supported by Shanghai Institutions of Higher Learning, the Science and Technology Commission of Shanghai Municipality (11DZ2260700) and by the US National Science Foundation, under the Grant No. PHY-1066278. We thank Chunjing Xie and Changbo Fu for very useful discussions.
}

\end{document}